\documentstyle[12pt]{article}
\newcommand{\unmezzo} {\frac{1}{2}}

\newcommand{\dt}{\partial _t}
\newcommand{\adv}[1]{(\vec{#1} \cdot \vec{\nabla})\vec{#1} }
\def\beq{\begin{equation}}
\def\eeq{\end{equation}}

\begin{document}

\title{ EXTENDED SELF SIMILARITY IN NUMERICAL SIMULATIONS OF 3D ANISOTROPIC
TURBULENCE.}
\author{ R. BENZI$^1$, M. V. STRUGLIA$^1$, R.
TRIPICCIONE$^2$
\\[1.5em]
$^1$ Dipartimento di Fisica,\\
Universit\`a di Roma {\it Tor Vergata},\\
Via della Ricerca Scientifica 1, 00133 Roma, Italy\\
and\\ Infn, Sezione di Roma {\it Tor Vergata}\\
$^2$ Infn, Sezione di Pisa, S. Piero a Grado, 50100 Pisa, Italy\\
\\
{\em Submitted to Phys. Rev. E \/} }
\maketitle

\begin{abstract}
Using a code based on the Lattice Boltzmann Equation, we have performed
numerical simulations of a turbulent shear flow.
We investigate the scaling behaviour of the structure
functions in presence of {\em anisotropic \/} homogeneous turbulence, and we
show that although
Extended Self Similarity does not hold when strong shear effects
are present, a more generalized scaling law can still be defined.
\end{abstract}

\vfill
{\bf PACS: 47.27.-i, 05.45.+b }
{\bf \hfill ROM2F/95/22 }
\newpage

In the last few years there has been a growing attention on the scaling
properties of {\em fully developed turbulence \/} and, in particular, on the
characterization of the Probability Distribution Function of the velocity
increments $\delta _r v\equiv v_x(\underline{x}+\underline{r})-v_x(\underline
{x})$, i.e. the velocity difference in the $ x-$direction  between two points
at distance $ r$.\\
To this aim, usually one considers  the scaling properties of the structure
functions defined as:
\beq
F_n(r)= \langle \vert \delta _r v\vert ^n \rangle.
\label{moments}
\eeq
According to the Kolmogorov theory\cite{K41} a scaling law for (\ref{moments})
is expected to hold in the so called inertial range, $ \eta\ll r\ll L$,( $ L$
being the integral scale  of the flow and $ \eta$ the Kolmogorov scale):
\beq
F_n(r) = A_n \left( \epsilon r \right)^{\frac{n}{3}}
\label{kolmo}
\eeq
where $ A_n$ are dimensionless constants and $ \epsilon $ is the mean rate of
energy dissipation.\\
There has been many experimental and numerical results suggesting that, because
of the {\em intermittency \/} of the velocity field, the relation (\ref{kolmo})
is violated, giving an {\em anomalous scaling law \/} with scaling exponents
$\zeta_n\neq\frac{n}{3}$.\\
By taking into account the fluctuations of the energy dissipation field, the
equation (\ref{kolmo}) has been modified by Kolmogorov\cite{K62} who introduced
the {\em Refined Similarity Hypothesis \/} (RSH):
\beq
F_n(r) = A_n \langle \epsilon_r ^{\frac{n}{3}}\rangle r^{
\frac{n}{3}}
\label{rksh}
\eeq
where $\epsilon_r $ is the local rate of energy transfer.\\
\[
\epsilon _r \equiv \frac{1}{r^3}\int _{B(r)} \epsilon (\underline{x}) d^3x.
\]
At present, most of the efforts, both theoretical and experimental, are devoted
to the determination of the anomalous scaling exponents and to the
investigation of the role played by the RSH.\\
The aim of this work is to investigate the scaling properties of the structure
functions in the case of a homogeneous shear flow, as a simple example of
anisotropic homogeneous turbulence. We are mainly interested to study the
scaling laws of the structure functions and to establish if the Extended Self
Similarity, recently introduced in literature\cite{ess1,ess3}, still holds for
shear flows, i.e. in presence of a non isotropic turbulent flow.\\
In this Letter we  first remind some concepts about Extended Self Similarity
(E.S.S.) and its relevance in order to estimate the $ \zeta_n$.\\
Next we briefly describe the shear flows and some of their properties. Finally
we discuss the numerical simulation and show that E.S.S does not hold for shear
flows, while a generalized scaling law, involving both E.S.S. and R.S.H. is
valid.\\
In principle, we can determine the scaling exponents $ \zeta _n$ by means of
experimental and numerical measures, but in the latter case some technical
problems arise.\\
 The highest Reynolds numbers that can be achieved by laboratory experiments
are about $10^{6\div 7} $, while the numerical simulations performed with the
most powerful computers now avalaible can reach $Re \sim 10^3$.
As the computational effort grows like $ Re ^3$, it could seem very hard to
obtain good estimates, at least comparable to the experimental results, of the
scaling exponents by the numerical simulations.\\
The concept of Extended Self Similarity (E.S.S.) can help us to fill up this
gap.\\
The idea is to investigate the scaling behaviour of one structure function
against the other, namely
\beq
F_n(r) \sim F_m(r)^{\beta(n,m)}
\label{ess}
\eeq
In particular it is expected that, at least in the inertial range,
$ \beta(n,3)= \zeta _n$.\\
Actually, there is strong evidence that E.S.S. is a powerful tool to
investigate the scaling laws and that it has many advantages
respect to the usual scaling against  $ r$, namely:\\
- it holds down to the dissipative range $ r\sim 4 \div 5 \eta$,\\
- it holds also for low Reynolds numbers.\\
Last but not least, the two previous properties allow a very accurate
determination of the scaling exponents. Indeed,
the $ \zeta _n$ can be estimated with an error of just a few percent.\\
The above statements can be summarized as follows.
We can always write the structure functions in the following way:
\beq
F_p(r)= C_p U_0^p \left[ \frac{r}{L} f_p\left(\frac{r}{\eta}\right)\right]
^{\zeta _p}
\label{essfp}
\eeq
with $ U_0^3 = F_3(r)$, $ L=U_0^3/\epsilon$ being the integral scale, and $
C_p$ dimensionless constants selected in such a way that $ f_p(r/\eta )=1$ for
$ r \gg \eta$.\\
E.S.S. implies that, for all the orders $ p$, the
function $ f_p(r/\eta )\equiv f(r/\eta)$ is the same.\\
We want to understand which are the effects of the lack of isotropy on the
anomalous scaling law defined in (\ref{ess}). To this effect,
we consider a simple shear flow.\\
Let us consider the usual Navier-Stokes equations  describing
a viscous, incompressible fluid of density $ \rho$, and velocity field
$ \vec{v}(\underline{x}, t)$:
\beq
\dt \vec{v} + \adv{v} = - \frac{1}{\rho} \vec{\nabla} p + \nu \Delta \vec{v}
+ \vec{f}
\label{NavSto}
\eeq
\[
\vec{\nabla} \cdot \vec{v} = 0
\]
Let us indicate the stationary solution of the above equations as $ \vec{U}$,
and define the turbulent velocities $ \vec{w}$ as:
\beq
\vec{v} = \vec{U}+\vec{w}.
\label{subst}
\eeq
In order to simplify the following discussion we choose the $ x-$direction
as  the direction of the main flow: $ U_x = U$, $ U_y = 0$, $ U_z=0$.\\
We have a {\em homogeneous shear flow \/}\cite{shear1} when the
main motion has a constant velocity in a given direction and a constant lateral
velocity gradient throughout the whole field, e.g. $U_x = U(z)$ and $\frac{d
U_x}{dz}= S$, so there is an evident lack of isotropy in the system.\\
Moreover we have a non zero turbulence shear stresses tensor, the component
$ \langle w_x w_z \rangle$ is different from zero and it makes a positive
contribution only to $ \dt \langle w_x^2\rangle$, resulting in non isotropy.\\
A generalization of the "$ \frac{4}{5}$" Kolmogorov equation for anisotropic
homogeneous shear flow \cite{Hinze} suggest that the typical scale fixed by
the shear intensity is $ r_s \sim (\epsilon / S^3)^{\unmezzo}$.
With zero shear this scale is infinite, otherwise it has a finite value: below
this scale the shear effects are expected to become negligible.\\
The particular question we want to address is: what does it happen to the
scaling laws (\ref{ess}) when  $ r_s$ falls into the inertial range?\\
In order to answer this question we perform a direct numerical simulation of a
turbulent shear flow, using a code based on the {\em Lattice Boltzmann
Equation\/}, for computational details see, for instance, \cite{LBE1,LBE2}.\\
We simulate a $ 3$D fluid occupying a volume of $V= L^3$ sites
with $ L=160$,  viscosity $ \nu =0.014$, and
obeying to the usual N-S equations plus a forcing term $ \vec{f} =(f_x(z), 0,
0)$ chosen such that the stationary solution of the N-S equations is:
\beq
U_x = A sin(k_z z) ~~~~~~~ U_y =0 ~~~~~~~ U_z=0.
\label{station}
\eeq
$ k_z= \frac{8\pi}{L} $ being the wave vector corresponding to the integral
scales, and $ A=0.3$.\\
In this way the shear has a spatial dependence $ S(z)\sim cos(k_z z)$. We can
access both  zones where the shear is maximum and locally homogeneous, and
 zones where the shear is minimum.\\
We evaluated $ v_{rms}$ as the mean value of $ (\frac{2}{3} E)^{1/2}$.
The simulations have been done at $ Re_\lambda = \frac{\lambda v_{rms}}
{\nu} \sim 40$, with $ \lambda \sim 15$ lattice spacings, and the Kolmogorov
scale is about $ 1$ lattice spacing wide.\\
The simulation has advanced $ 100 000$ iterations corresponding to about $ 25$
macroscale eddy turnover times $ \tau _0 \sim L/v_{rms}$: $ 40$ velocity
configurations have been saved
every $ 2500$ time steps, in order to ensure the statistical independence of
the
different configurations.\\
We have evaluated the structure functions $ F_n(r)$ up to the tenth order.
The mean values of $ \vert \delta _r v\vert ^n$ have been evaluated through
time and spatial average at fixed z-level:
\[
\langle O(\underline{r},t) \rangle = \frac{1}{T} \int_0^T dt~ \frac{1}{L^2}
\int
dx dy~ O(\underline{r},t).
\]
In Fig.1 we have a log-log plot of the longitudinal ($
x$-direction) structure function $ F_6(r)$ against $ F_3(r)$, obtained from
the velocity fields corresponding to the minimum shear level.
The dashed curve is the best fit done in the range between the $ 20$-th and $
30
$-th grid point, and corresponds to a slope of $ 1.79$  in good agreement with
other measured  values of $ \zeta _6$.
Every point in the plot corresponds to a grid point and the lattice
spacing is $ \sim 1 \eta$ wide. As we can see the E.S.S. holds as usual until $
4\div 5 \eta$.\\
Fig.2 shows the same plot but at the maximum shear level. It is
quite evident that E.S.S. does not hold. In any case,
the slope corresponding to
the best fit can be estimated at about $ 1.43$, quite different
from the previous value.
Similar results have been obtained for all the others structure
functions.\\
 \begin{table}[htp]
 \begin{center}
 \begin{tabular}{|r|r|r|r|r|r|r|}
 \hline
 \hline
       & $ $
       & $ \zeta _2$
       & $ \zeta _4$
       & $ \zeta _6$
       & $ \zeta _8$
       & $\zeta _{10}$
 \\
 \hline
       & min sh
       & $ 0.70$
       & $1.28$
       & $1.79$
       & $2.25$
       & $2.68$
 \\
\hline
       & max sh
       & $ 0.76$
       & $1.18$
       & $1.43$
       & $1.56$
       & $1.61$
 \\
\hline
       & SL mod.
       & $ 0.696$
       & $1.279$
       & $1.778$
       & $2.211$
       & $2.593$
 \\
\hline
\hline
\end{tabular}
\caption[]{Scaling exponents evaluated at the minimum shear (first line), at
the maximum shear (second line), and from the She-Leveque\cite{SL} model.}
 \end{center}
 \label{tab}
 \end{table}
In Table 1 \ref{tab}  we show the scaling exponents obtained for the even
order structure functions.\\
We can suggest the following explanation for the different scaling behaviour
in presence of shear.
In our simulations the scale  $ r_s$ is about $ 4$ lattice spacings at the
maximum shear level, so the entire range over which the E.S.S. holds (see
Fig.1)
is subjected to the shear effects.\\
 Our result clearly shows that the shear completely destroys the E.S.S.\\
We now turn our attention to RSH. Following \cite{ess3} we can consider the
generalization of RSH by introducing an {\em effective scale \/} $S(r)\equiv
\langle \delta_r v^3\rangle/\langle \epsilon_r\rangle= r f(r/\eta)$.
Then ESS combined with RSH suggests:
\beq
\frac{\delta _r v ^3}{S(r)} \sim \epsilon _r
\label{refKhyp}
\eeq
If the equation (\ref{refKhyp}) is true, as it has already been verified for
experimental data sets referring to homogeneous and isotropic
turbulence\cite{ess3}, we expect that the local rate of energy
transfer and the structure functions satisfy the following scaling law:
\beq
\langle \delta _r v ^{3n}\rangle\sim \langle \epsilon_r^n\rangle \langle
\delta_r v ^3\rangle ^n
\label{sesscal}
\eeq
over a range wider than the inertial one.\\
Using the data from our simulation, we obtained the results shown in
Fig.s 3-4.\\
As we can see the scaling of $ \langle \epsilon ^2\rangle \langle \vert \delta
_r v\vert ^3\rangle ^2$ against $ \langle  \delta _r v ^6\rangle$ is well
verified in both the zones of maximum and minimum shear with a slope very close
to one. This result is extremely interesting and suggests that the scaling law
(\ref{sesscal}) is universal, regardless the isotropy conditions of the
turbulent flow.\\
Let us summarize the results that have been obtained and suggest a possible
interpretation for them and what should be their future developments.\\
First of all it has been shown that E.S.S. does not hold for anisotropic
turbulent flows, according to similar results obtained from experimental data
sets of turbulent boundary layers \cite{Sreeniv}, where strong shear effects
are expected to appear.\\
It means that moments of different order show a different dependence from the
cutoff scale. This means that the shear affects the function
$f_p(r/\eta)$, defined in (\ref{essfp}), which is no longer the same for all
the orders $p$.\\
Nevertheless, the scaling law (\ref{sesscal}) is valid even in presence of
shear
and at the smallest scales investigated, suggesting that the scaling law of
a generic structure function is related to those of the  third one and of the
energy dissipation in a universal way, for all analyzed scales, a
remarkably non-trivial result.\\
We think that the investigation of the self-scaling properties of the energy
dissipation $ \epsilon_r$ would deserve more attention,  in order to
understand how the structure functions of the velocity increments depend on the
resolution scale and to explain the ESS violation  in shear flows.\\
A deeper analysis of these arguments, together with other numerical and
experimental results, will be the subject for further investigation
\cite{BBCST,long}.\\

We thank L.Biferale for the interesting discussion we have had.
M.V.S. acknowledge also  F. Massaioli, S. Succi, and A. Vicer\`e
for their useful advice about the LBE code and the parallel computer APE that
has been used to run it.\\
This work was partially supported by the EEC contract CT93-EV5V-0259.\\

\vfill
\newpage
{\bf Figure Captions.}
\begin{description}
\item[Figure 1.]
Log-log plot of $ F_6(r)$ against $ F_3(r)$ at the minimum shear. The dashed
line is the best fit with slope $ 1.79$. Every point in the plot corresponds
to a grid point and the lattice spacing is $ \sim 1 \eta$ wide.
\item[Figure 2.]
The same as in Fig.1 at the maximum shear. The dashed
line is the best fit with slope $ 1.43$.
\item[Figure 3.]
Plot of $ \langle \epsilon ^2\rangle \langle \vert \delta
_r v\vert ^3\rangle ^2$ against $ \langle  \delta _r v ^6\rangle$ at the
 minimum shear. The points refer to the scales at $2,4,5,8,10,16,20,32,
 40$ grid points and the dashed line is the best fit done over these points,
 corresponding to the slope $ 0.99$.
\item[Figure 4.]
The same as in Fig.3 at the maximum shear. The dashed
line is the best fit with slope $ 0.99$.
\end{description}
\vfill
\end{document}